\documentclass[9pt,twocolumn,twoside]{pnas-new}

\templatetype{pnasresearcharticle} 

\usepackage{units}
\usepackage{tabularx}
\usepackage{multirow} 
\usepackage{makecell} 
\usepackage{booktabs} 

\usepackage{subcaption}

\usepackage[numbers,sort&compress]{natbib}  
  
\usepackage[capitalize,nameinlink]{cleveref} 

\title{In vivo investigation of the multi-exponential T$_\mathbf{2}$ decay in human white matter at 7 T: Implications for myelin water imaging at UHF}

\author[1,2,3]{Vanessa Wiggermann}
\author[1,3,4]{Alex MacKay}
\author[1,2,3,5,*]{Alexander Rauscher}
\author[6,*]{Gunther Helms}

\affil[1]{Department of Physics and Astronomy, University of British Columbia}
\affil[2]{Department of Pediatrics, University of British Columbia}
\affil[3]{UBC MRI Research Centre, University of British Columbia}
\affil[4]{Department of Radiology, University of British Columbia}
\affil[5]{BC Children's Hospital Research Institute, University of British Columbia}
\affil[6]{Department of Clinical Sciences Lund, Lund University}
\affil[*]{equal contribution}

\leadauthor{Wiggermann} 

\authordeclaration{The authors have no conflicts of interest to
  declare.}
\correspondingauthor{\textsuperscript{a}Corresponding author. E-mail: vanessaw@drcmr.dk}


\begin{abstract} 
Multi-component T$_\mathbf{2}$-mapping using a gradient and spin-echo (GraSE) acquisition has become standard for myelin water imaging at 3T. Higher magnetic field strengths promise signal-to-noise ratio benefits but face specific absorption rate limits and shortened T$_\mathbf{2}$ times. This study investigates compartmental T$_\mathbf{2}$ times in vivo and addresses advantages and challenges of multi-component T$_\mathbf{2}$-mapping at 7T.
We acquired 3D multi-echo GraSE data in seven healthy adults at 7T, with three subjects scanned also at 3T. Stimulated echoes arising from B$_\mathbf{1}^\mathbf{+}$-inhomogeneities were accounted for by the extended phase graph (EPG) algorithm. We used the computed T$_\mathbf{2}$ distributions to determine T$_\mathbf{2}$ times that identify different water pools and assessed signal-to-noise and fit-to-noise characteristics of the signal estimation. We compared short T$_\mathbf{2}$ fractions and T$_\mathbf{2}$ properties of the intermediate water pool at 3T and 7T. 
Flip angle mapping confirmed that EPG accurately determined the larger B$_\mathbf{1}^\mathbf{+}$-inhomogeneity at 7T. Multi-component T$_\mathbf{2}$ analysis demonstrated shortened T$_\mathbf{2}$ times at 7T compared to 3T. Fit-to-noise and signal-to-noise ratios were improved at 7T but depended on B$_\mathbf{1}^\mathbf{+}$-homogeneity. Lowering the shortest T$_\mathbf{2}$ to 8 ms and adjusting the T$_\mathbf{2}$ threshold that separates different water compartments to 20 ms, yielded short T$_\mathbf{2}$ fractions at 7T that conformed to 3T data. Short T$_\mathbf{2}$ fractions in myelin-rich white matter regions were lower at 7T than at 3T, and higher in iron-rich structures. 
Adjusting the T$_\mathbf{2}$ compartment boundaries was required due to the shorter T$_\mathbf{2}$ relaxation times at 7T. Shorter echo spacing would better sample the fast decaying signal but would increase peripheral nerve stimulation. Multi-channel transmission will improve T$_\mathbf{2}$ measurements at 7T. 
We used a multi-echo 3D-GraSE sequence to characterize the multi-exponential T$_\mathbf{2}$ decay at 7T. We adapted T$_\mathbf{2}$ parameters for evaluation of the short T$_\mathbf{2}$ fraction. Obtained 7T multi-component T$_\mathbf{2}$-maps were in good agreement with 3T data. 
\end{abstract}

\dates{This is the pre-peer reviewed version of the following article: Wiggermann et al.\ NMR Biomed 2020, which has been published in final form at  DOI:10.1002/nbm.4429. This article may be used for non-commercial purposes in accordance with Wiley Terms and Conditions for Use of Self-Archived Versions.}
\doi{bioR${{\chi}}$iv}

\begin{document}

\maketitle
\thispagestyle{firststyle}
\ifthenelse{\boolean{shortarticle}}{\ifthenelse{\boolean{singlecolumn}}{\abscontentformatted}{\abscontent}}{}



\dropcap{T}he ability to non-invasively obtain surrogate metrics for myelin concentrations in vivo, makes MRI ideally suited for studying brain myelination during early development \cite{1,2}, alterations in myelin during healthy aging \cite{3}, concussion \cite{4} and in neurodegenerative diseases, such as multiple sclerosis (MS) \cite{5}. The increased availability of 7 T MR scanners generates the need for transferring myelin-specific techniques from 3 T to ultra-high fields (UHFs). T$_2$-based myelin water imaging (MWI), in particular, has proven to be specific to myelin lipids \cite{6,7,8} and less sensitive to axons and variations in T$_1$ \cite{9} than other myelin-sensitive techniques \cite{10,11}. Axonal myelin, which ensures rapid neural signal transmission, consists of multiple phospholipid bilayer sheaths. Water trapped within these bilayers, the myelin water (MW), has restricted mobility and thus exhibits faster T$_2$ decay than water in intra- and extracellular spaces (IEW). In MWI, the short T$_2$ signal fraction is extracted from measured multi-exponential T$_2$ decay data, which are characteristic for myelinated tissues of the central and peripheral nervous system \cite{12,13}.

To date, MWI has been performed at field strengths up to 3 T. Given that the MW fraction computation is sensitive to noise \cite{14}, MWI may benefit from the increased signal-to-noise ratio (SNR) at 7 T. However, increased B$_1^+$-inhomogeneity and specific absorption rate (SAR) \cite{15} challenge its implementation at UHFs. With the introduction of the extended phase graph (EPG) \cite{16,17}, the effect of B$_1^+$-inhomogeneity on the T$_2$ decay can now be accurately modeled. Furthermore,whole brain 3D encoding has become possible by replacing the Carr-Purcell-Meiboom-Gill (CPMG) with an accelerated gradient- and spin-echo (GraSE) acquisition \cite{18}, which is now the standard at 3 T. Prolonged data acquisition times at UHFs that result from longer TR to meet SAR constraints can also be alleviated by GraSE acceleration.

Here, we adapted a multi-echo 3D GraSE protocol from 3 T to 7 T to investigate the feasibility and challenges of estimating the myelin-associated short T$_2$ fraction in vivo at 7 T. Because changes in T$_2$ times occur \cite{19}, it is critical to determine the 7 T T$_2$ characteristics of all water pools for in vivo MWI \cite{20}. We performed multi-component T$_2$ analysis of the acquired decay data, including application of the EPG algorithm \cite{16,17} to account for stimulated echoes due to imperfect signal refocusing. We then studied the T$_2$ distributions in healthy white matter (WM) at 7 T and assessed the noise properties of the data. T$_2$ component boundaries appropriate for 7 T were established based on data from three travelling brains. Estimated short T$_2$ fractions in white and gray matter regions at 7 T were compared with those at 3 T. 

\section*{Experimental}
\subsection*{In vivo data acquisition}
Seven healthy adults (5 male/ 2 female, age range 21 to 51 years, mean 29 years) were examined on an actively shielded 7 T whole-body Achieva MR system (software level R5.1, Philips Healthcare, Best, The Netherlands, located at the National 7 T facility in Lund, Sweden), using a dual transmit birdcage coil and a 32-channel receive array (Nova Medical, Boston, MA). For comparison, three of these subjects were also scanned on a 3 T Philips Achieva MR system at the University of British Columbia, Vancouver, Canada, using the body transmit coil and an 8-channel receive head coil. Examinations had been approved by the local ethics committees at both sites and written informed consent was obtained prior to each session.

In line with the current implementation of MWI at 3 T$_1$ \cite{18}, an accelerated 3D 32-echo GraSE acquisition (EPI factor = 3) was used at both field strengths. Acquisition parameters were: TE$_1$/$\Delta$TE = 10 \unit{ms}/10 ms, acquired voxel size $0.98\times0.98\times5$ mm$^{3}$ with a slice oversampling factor of 1.3, reconstructed to $0.98\times0.98\times2.5$ mm$^{3}$, 40 transversal slices. The 3D volume was aligned to the subcallosal plane. TR was 1000 ms at 3 T and 1670 ms at 7 T, the minimum TR compliant with SAR limits. The latter value was obtained using non-selective radiofrequency pulses of the smallest permitted amplitude, 10 $\mu$T. Over-contiguous slices were acquired and a SENSE factor of two was used in the left-right direction at both field strengths. At 7 T, the short axial extent the head coil allowed for additional SENSE acceleration in the head-feet direction by a factor of 1.5, yielding a total acquisition time of 11 min per scan as compared to 14 min at 3 T. In addition, a 3D turbo field echo T$_1$-weighted anatomical reference of 200 sagittal images was acquired with TI = 1200 ms, 3500 ms shot interval, $0.9\times0.9\times0.9$ mm$^{3}$ voxel size, readout TE/TR = 1.97/8 ms, flip angle (FA) = 8$^{\circ}$, SENSE = 2.5 in slice direction. Axial FA maps were acquired using DREAM \cite{21} at TE$_{\text{STE}}$/TE$_{\text{FID}}$ = 1.81 \unit{ms}/1.4 \unit{ms}, preparation FA = 40$^{\circ}$ and read-out FA = 7$^{\circ}$.

\subsection*{GraSE data processing}

All data analysis was conducted at the University of British Columbia using in-house developed software running on MATLAB 2017b. A detailed description of the T$_2$ analysis can be found elsewhere \cite{20,22}. Briefly, EPG was employed to model stimulated echoes in the T$_2$ decay arising from imperfect refocusing angles in the echo train \cite{16,17}. Thereafter, T$_2$ distributions were extracted on a voxel-by-voxel basis by regularized non-negative least squares fitting (T$_2$ grid with 40 T$_2$ times, $\chi^{2}$/$\chi^{2}_{min}$ = 1.02). Different water compartments of a voxel were distinguished by their specific T$_2$ intervals. The short fraction signifies signal contributions between the shortest T$_2$ time (T$_{2,min}$) and the T$_2$ cut-off time (T$_{2,cut}$), separating short T$_2$ signal from other, intermediate T$_2$ signal components. Adequate T$_2,min$ and T${_2,cut}$ were established by comparison of 7 T to 3 T data. By choosing the maximum T$_2$ = 2s, we accounted for occasional long T$_2$ signal originating from cerebrospinal fluid. T$_2$ signal fractions were estimated from the integral of the respectively established T$_2$ interval divided by the integral of the entire T$_2$ distribution \cite{12}. In the following, we refer to short and intermediate T$_2$ fractions, peaks and water pools, rather than explicitly calling them MW and IEW. In addition to estimating the pool fractions, the geometric mean T$_2$ of the intermediate water pool, EPG-derived FA maps, as well as the signal-to-noise (SNR) and fit-to-noise ratio (FNR) per voxel were computed. The surrogate measure of SNR is the ratio of the signal from the first echo, relative to the square root of the variance of the fitting residuals, i.e. the difference between the computed and measured T$_2$ decay. The FNR is the sum of all fitted T$_2$ signal amplitudes relative to the square root of the variance of the fitting residuals. These parameters were assessed to establish goodness-of-fit and reliability of the short T$_2$ estimation.

\subsection*{Region-of-interest (ROI) analysis}
Quantitative ROI analysis was performed in each subject’s GraSE-space. The average short T$_2$ fractions, the geometric meanT$_2$ values and FNR were evaluated within seven ROIs. In line with literature comparing MWI at 1.5 T and 3 T \cite{23,24,25}, the ROIs comprised four WM regions: the genu and splenium of the corpus callosum (CC), the internal capsules (anterior and posterior, IC) as well as the superior longitudinal fasciculus, and three regions of gray matter (GM): the thalamus, putamen and cingulate gyrus. To obtain the ROIs, we reoriented the sagittal 3D T$_1$-weighted volumes to axial slices, corrected for intensity biases by the N3-algorithm as implemented in FreeSurfer \cite{26}, and performed brain extraction in FSL \cite{27}. The T$_1$-weighted volumes were then non-linearly registered to the MNI152 standard space using FLIRT \cite{28,29} and FNIRT \cite{30,31}. Subsequently, the matrices were inverted to warp the Harvard-Oxford subcortical 1 mm atlas \cite{32} and the Jülich ICBM-1mm labels \cite{33} onto the individual T$_1$-spaces. The T$_1$-weighted images and the atlas-based ROIs were then mapped onto a single-echo of the 3D-GraSE sequence using NiftyReg \cite{34}. ROI voxels affected by interpolation and cerebrospinal fluid were removed by thresholding. To compare the 7 T and 3 T data in the travelling brains, the 7 T GraSE ROIs were linearly co-registered to the 3 T GraSE data using NiftyReg \cite{34}.

\section*{Results}
\subsection*{Comparison of individual 3 T and 7 T GraSE images}
\begin{figure}
\centering
\includegraphics[width=1.0\linewidth]{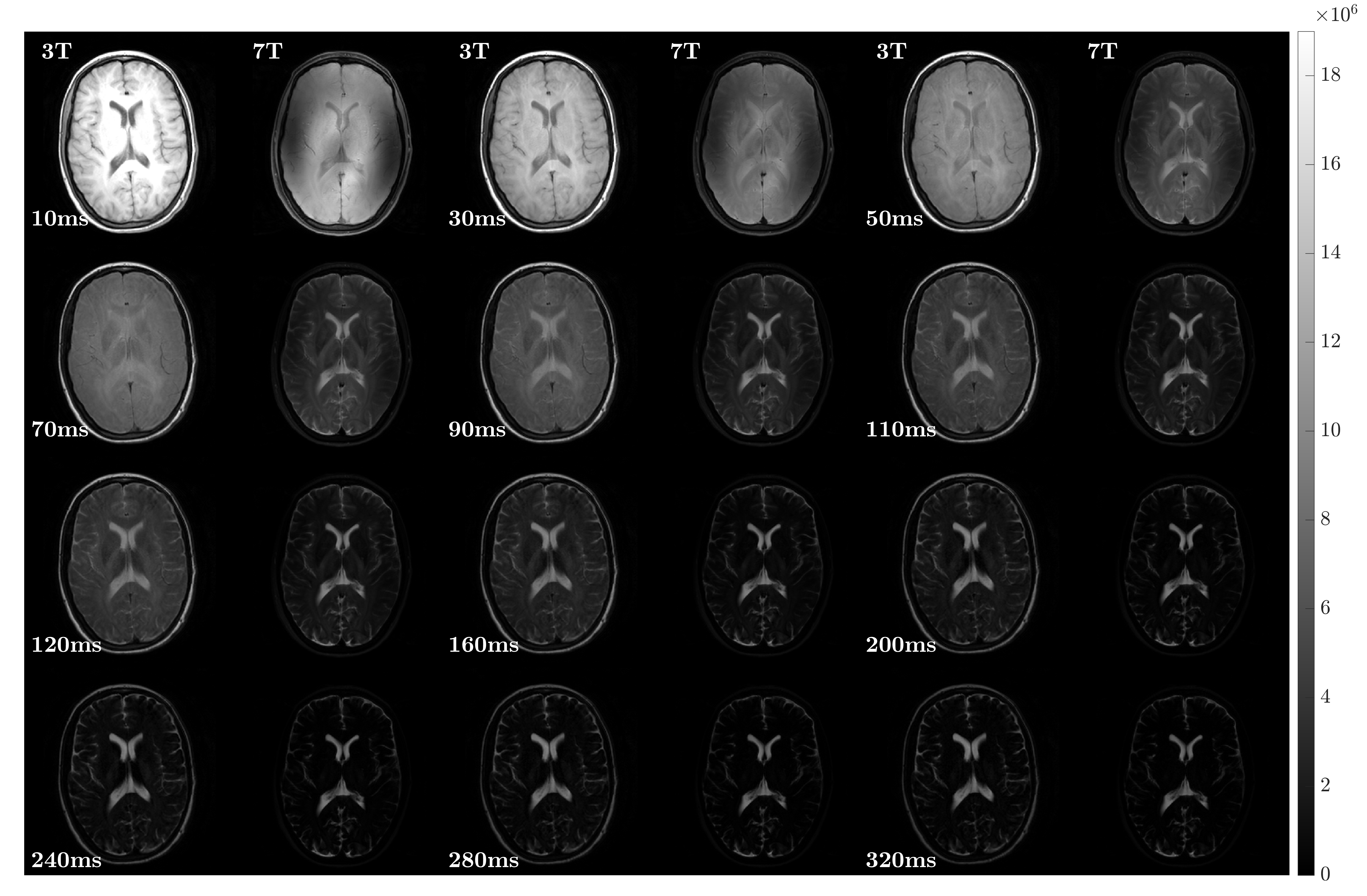}
\caption{\textbf{GraSE-images at 3 T and 7 T of one volunteer showing greater B$_1^+$-inhomogeneity, reduced T$_1$-weighting and a more rapid T$_2$ decay at 7 T.} Every second echo is depicted up to echo 12 and every 4th echo thereafter. Corresponding TEs are reported in the bottom left corner of the respective images. Columns alternate 3 T and 7 T data. The scaling of the physical signal (in arbitrary units, intensity bar on the right) differed between the scanners.}
\label{fig:fig1}
\end{figure}

The TE dependence of unprocessed 3 T and 7 T GraSE images is shown in \cref{fig:fig1}. Images at 7 T exhibited faster signal decay than the 3 T images. Signal loss was most prominent in the temporal lobes due to low effective B$_1^+$. The first echo also demonstrated that T$_1$-weighting was much smaller at 7 T compared to 3 T, indicating that the SAR-related prolongation in TR compensated for longer T$_1$ at 7 T. 

\subsection*{Flip angle estimation by EPG}
\begin{figure}
	\centering
	\includegraphics[width=1.0\linewidth]{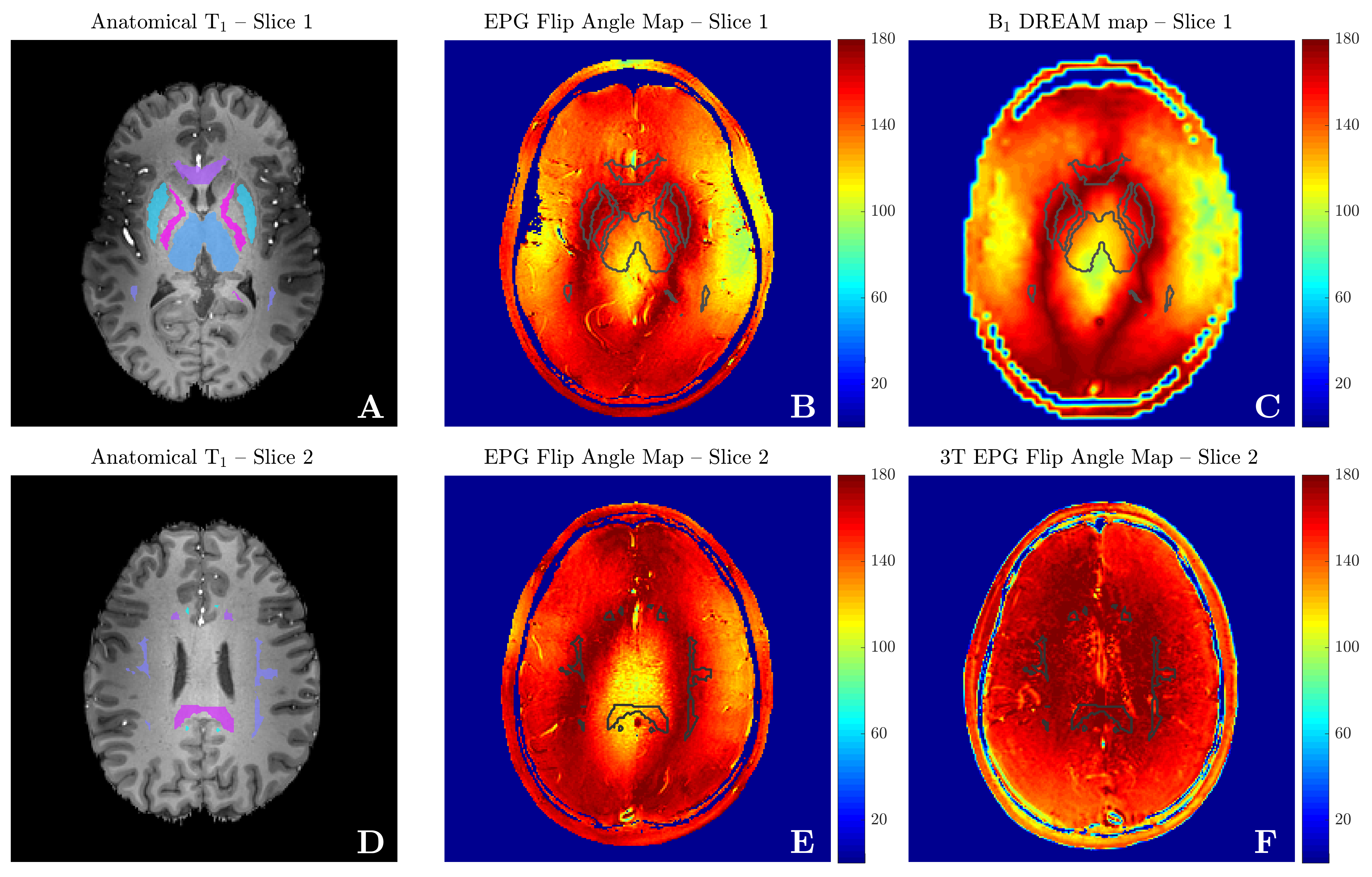}
	\caption{\textbf{FA maps at 7 T, contrasting EPG estimates with B$_1^+$ mapping, and 7 T in comparison to 3 T maps.} \underline{Top row}: ROIs of the genu of the corpus callosum (CC), the internal capsules (IC), putamen and thalamus are outlined on the FA maps and full ROIs are shown on the co-registered structural T$_1$w (A). Comparison of FA maps obtained by EPG (B) and DREAM B$_1^+$ mapping (C) at 7 T demonstrated good agreement, including in lateral regions with strongly reduced B$_1^+$. Difference maps of these two FA estimates are provided for the two image slices in (C) and (F). \underline{Bottom row}: Slice traversing the splenium of the CC and the superior longitudinal fasciculus (D). Greater B$_1^+$-variations at 7 T (E) are contrasted with the relatively more homogeneous FA distribution at 3 T (F). Note that actual angles of 180$^\circ$ + $\theta$ ($\theta$ > 0) are displayed as 180$^\circ$ – $\theta$ in the FA maps.}
	\label{fig:fig2}
\end{figure}

\cref{fig:fig2} shows anatomical and FA maps for two representative slices traversing the selected ROIs. EPG-estimates of refocusing FAs at 7 T agreed well with the independently acquired DREAM map (top row). The B$_1^+$-map was converted to refocusing FA $\leq$ 180$^\circ$ to match the output range of the EPG algorithm. Note the correspondence across the full B$_1^+$ range from the 50\% lateral drop-off to the 50\% increased B$_1^+$ in the centre of the brain. Thus, B$_1^+$-inhomogeneity throughout the brain at 7 T, including FAs of 140$^\circ$ – 240$^\circ$, was well accounted for when computing the individual T$_2$ distributions. In most ROIs, the deviation from 180$^\circ$ was moderate. In the bottom row, EPG-estimated refocusing FA maps at 3 T and 7 T are compared. At 3 T, FAs were consistently close to 180$^\circ$ around the splenium. By contrast, EPG-FAs at 7 T in the same area ranged between 120$^\circ$ – 180$^\circ$ (180$^\circ$– 240$^\circ$ true FAs). 

\subsection*{Adapting T$_{\text{2,min}}$ and T$_{\text{2,cut}}$  for T$_2$ distributions at 7 T}
For healthy human brain WM at 3 T, it has been described that the short T$_2$ fraction is encompassed by T$_{\text{2,min}}$ = 15 ms and T$_{\text{2,cut}}$ = 40 ms, and the intermediate T$_2$ component lays within 40 – 200 ms \cite{23,25}. The faster decay observed at 7 T suggests that shortening both T$_{\text{2,min}}$, the lowest boundary of the T$_2$ grid, and T$_{\text{2,cut}}$, the threshold differentiating the two water pools, will be required to render short T$_2$ fraction estimates consistent with 3 T. 
Using the 32-echo GraSE data of TE/$\Delta$TE = 10/10 ms at 3 T, a T$_{\text{2,min}}$ = 10 ms was sufficient to encompass the short T$_2$ signal in WM, which was located at approximately 15 ms (see \cref{sfig:fig1}). However, at 7 T, where the short T$_2$ peak is located at approximately 11 -- 12 ms, a T$_{\text{2,min}}$ of 8 ms may be required to fully capture the short T$_2$ signal. This selection of T$_{\text{2,min}}$ agrees with recent simulation work, recommending T$_{\text{2,min}}$ = TE$_1$ – 2 ms \cite{20}. To determine T$_{\text{2,cut}}$ at 7 T, we compared in Figure 3A the 7 T T$_2$ distribution across the splenium ROI for all volunteers. These distributions are averages of the single-voxel distributions. One can identify well separated modes of short and intermediate T$_2$ in all volunteers despite some variation in the intermediate T$_2$ peak, which comprised T$_2$ times up to 150 ms at 7 T. In the bottom panel (B), the average 7 T T$_2$ distribution of all subjects is compared with the average 3 T distribution, both for a logarithmically-spaced T$_2$ grid of 8 ms to 2 s. The curves indicate that the T$_2$ cut-off for 7 T should be on average about 20 ms to separate the different water peaks in all individuals. At 3 T, water pools seem best separated by choosing T$_{\text{2,cut}}$ = 30ms (see vertical lines in panel B). T$_2$ values of the intermediate peak ranged up to 200 ms at 3 T (B), in line with previous reports \cite{23}.

\begin{figure}
	\centering
	\begin{subfigure}[c]{\linewidth}
	\includegraphics[width=1.0\linewidth]{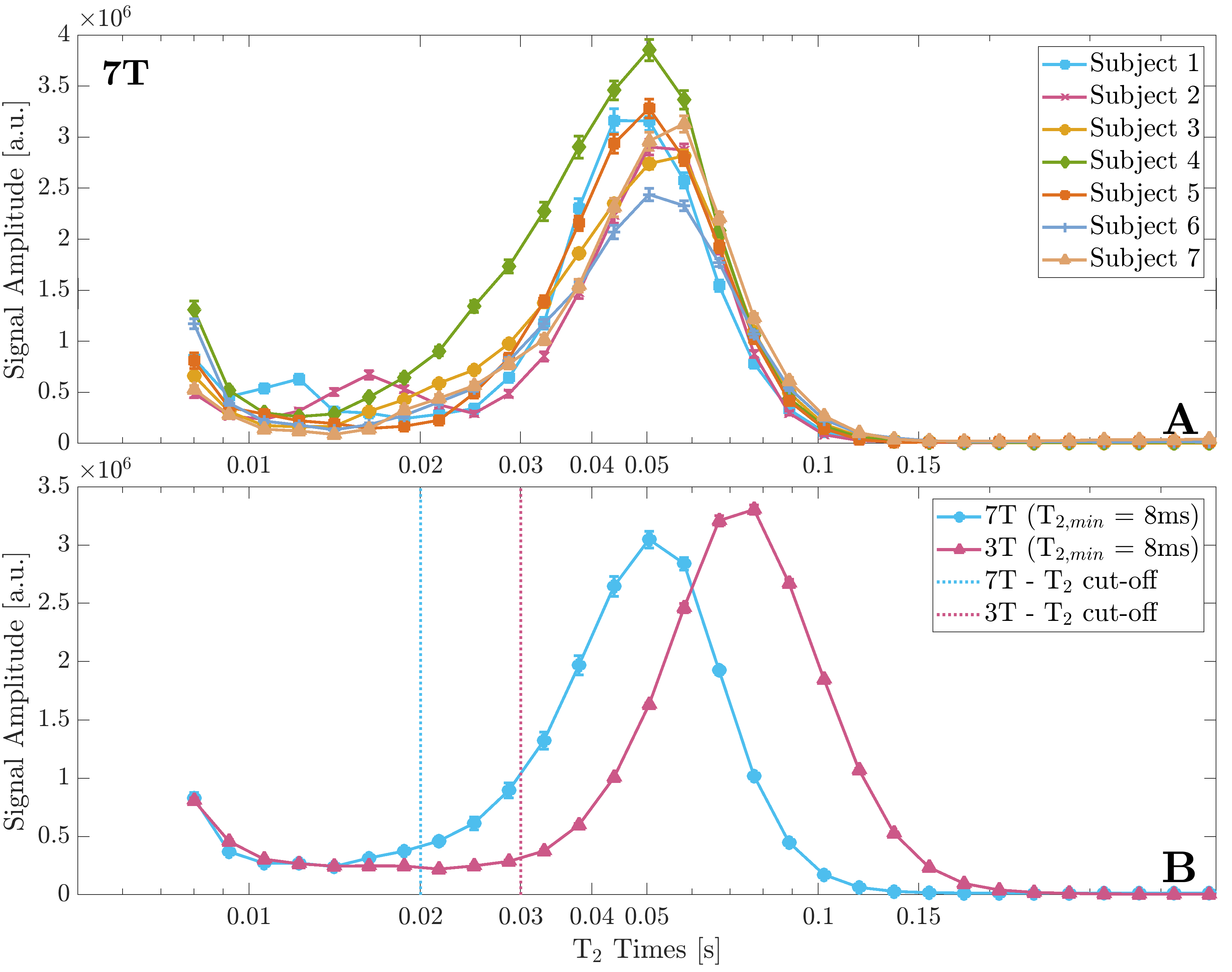}
	\phantomcaption
	\label{fig:fig3A}
	\end{subfigure}
	\begin{subfigure}[c]{0\linewidth}
	\includegraphics[width=0\linewidth]{Upload_Fig3}
	\phantomcaption
	\label{fig:fig3B}
	\end{subfigure}
    \vspace{-0.9cm}
	\caption{\textbf{Mean T$_2$ distribution at 7 T of the splenium shown for all seven subjects (top) and the subject average distributions (bottom) at 7 T (blue, circles) and 3 T (pink, triangles).} Separate short and intermediate T$_2$ contributions were detected in each subject, although the mean T$_2$ time of the short peak varied (A). Nevertheless, it is possible to compromise on a common T$_2$ cut-off (B). For the subsequent analysis, T$_{\text{2,cut}}$ was chosen to be 20 ms at 7 T and 30 ms at 3 T, as indicated by the vertical lines in B.}
	\label{fig:fig3}
\end{figure}

\begin{figure}
	\centering
	\includegraphics[width=1.0\linewidth]{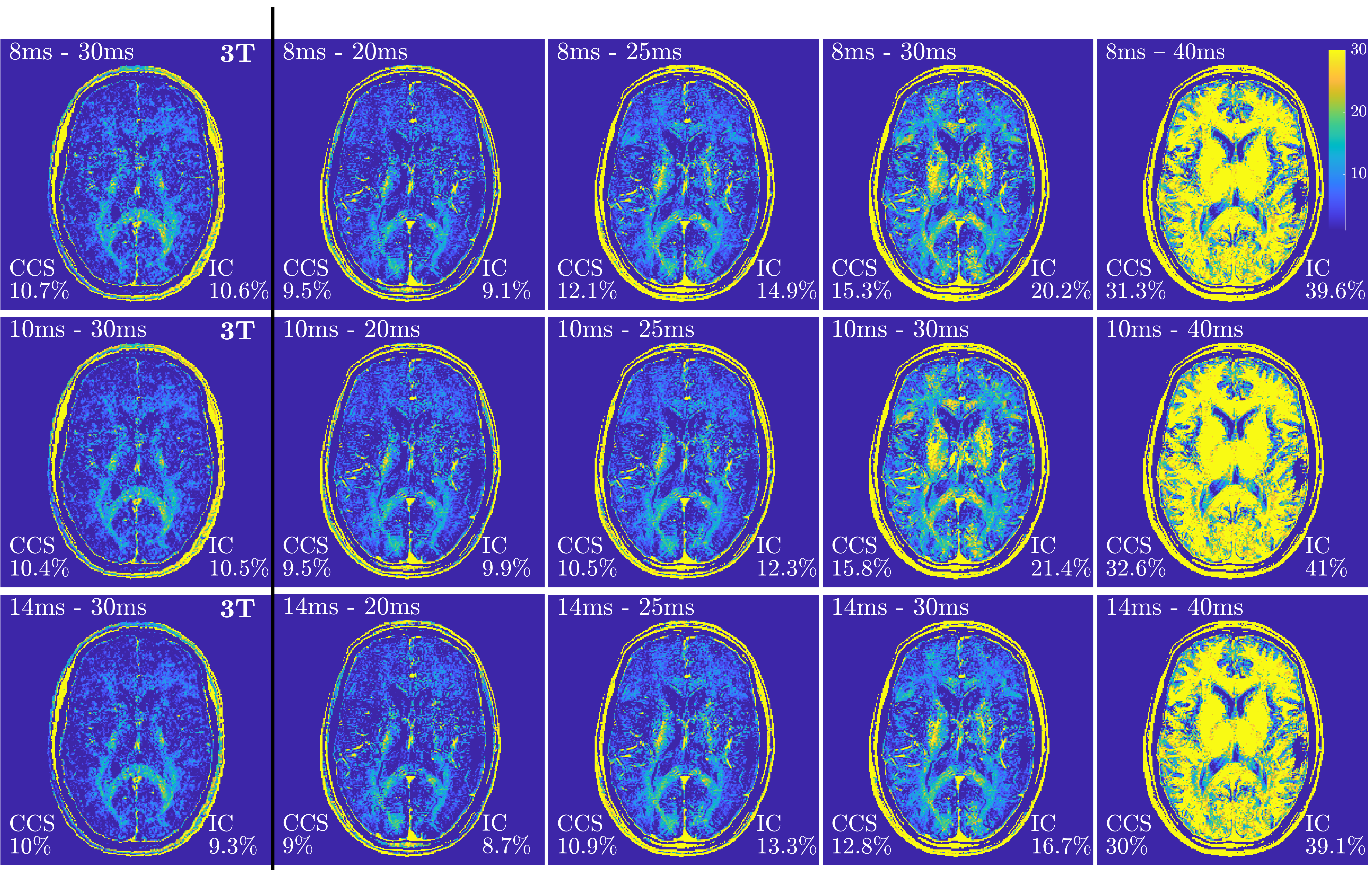}
	\caption{\textbf{Short T$_2$ fraction maps at different T$_{\text{2,cut}}$ (columns) and T$_{\text{2,min}}$ (rows).} Column 1 displays the reference 3 T data, computed with a T$_{\text{2,cut}}$ = 30 ms. Columns 2–-5 show the corresponding 7 T data. The respective T$_2$ ranges are given in each panel’s top left corner. The short T$_2$ fractions of the splenium of the CC (CCS) and the internal capsules (IC) are indicated in the bottom corners. A reduction of T$_{\text{2,min}}$ (bottom to top) and an increase of T$_{\text{2,cut}}$ (left to right) lead to increases in the corresponding short T$_2$ fractions, albeit due to different mechanisms. As evident from \cref{fig:fig3,fig:fig4}, with T$_{\text{2,cut}}$ above 20 ms, the short T$_2$ pool includes signal from the intermediate water peak. }
	\label{fig:fig4}
\end{figure}

Further inspection of 7 T T$_2$ distributions in the other WM regions confirmed that T$_{\text{2,min}}$ = 8 ms and T$_{\text{2,cut}}$ = 20 ms were generally applicable in WM voxels. This is affirmed in \cref{fig:fig4}, which depicts the impact of the selection of appropriate T$_{\text{2,min}}$ and T$_{\text{2,cut}}$ for the computation of short T$_2$ fraction maps. T$_{\text{2,min}}$ was varied through 8, 10 and 14 ms (rows), while choosing different T$_{\text{2,cut}}$ between 20, 25, 30 and 40 ms for the 7 T data (columns 2–5). The resulting images are compared to 3 T maps obtained at short T$_2$ intervals of 8 -- 30 ms, 10 -- 30 ms and 14 – 30 ms (column 1). The 7T short T$_2$ fraction maps that closest resembled the 3 T maps were those with T$_{\text{2,min}}$ of 8 ms or 10 ms and T$_{\text{2,cut}}$ of 20 ms or 25 ms. At the chosen cut-off T$_2$ times, the 8 -- 20 ms versus the 10 -- 20 ms maps at 7 T and the 8 -- 30 ms versus 10 -- 30 ms maps at 3 T did hardly yield noticeable differences for any of the outcome measures, i.e. short T$_2$ fraction, short and intermediate geometric mean T$_2$, for the subject of \cref{fig:fig4} (\cref{stab:tab1}). Changing T$_{\text{2,min}}$ for a given T$_{\text{2,cut}}$ resulted in minor variations in the short T$_2$ signal fraction of less than 1\%. Note that a 25 ms cut-off may yield partial signal misassignment from the intermediate peak to the short T$_2$ component window. This is more obvious at higher T$_2$ cut-offs, especially at T$_{\text{2,cut}}$ = 40 ms, where it resulted in widespread and excessively high short T$_2$ fractions of more than 30\%. Iron-rich deep brain nuclei, for instance the putamen, were emphasized in the 7 T maps when choosing T$_{\text{2,cut}}$ = 30 ms or higher (column 4, \cref{fig:fig4}). Larger veins were also more pronounced due to the shorter T$_2$ of deoxygenated blood at 7 T \cite{35}. 

For all further analyses, we delimited the short T$_2$ range to 8 -- 30 ms for 3 T and 8 -- 20 ms for 7 T. This range selection is supported by our observations in \cref{fig:fig3B} in addition to a small improvement in the FNR when comparing to T$_{\text{2,min}}$ = 10 ms (\cref{stab:tab1}).

\subsection*{Regional analysis of intermediate geometric mean T$_2$ and short T$_2$ fractions at 3 T and 7 T} 
\begin{figure}
	\centering
	\begin{subfigure}[c]{\linewidth}
	\includegraphics[width=1.0\linewidth]{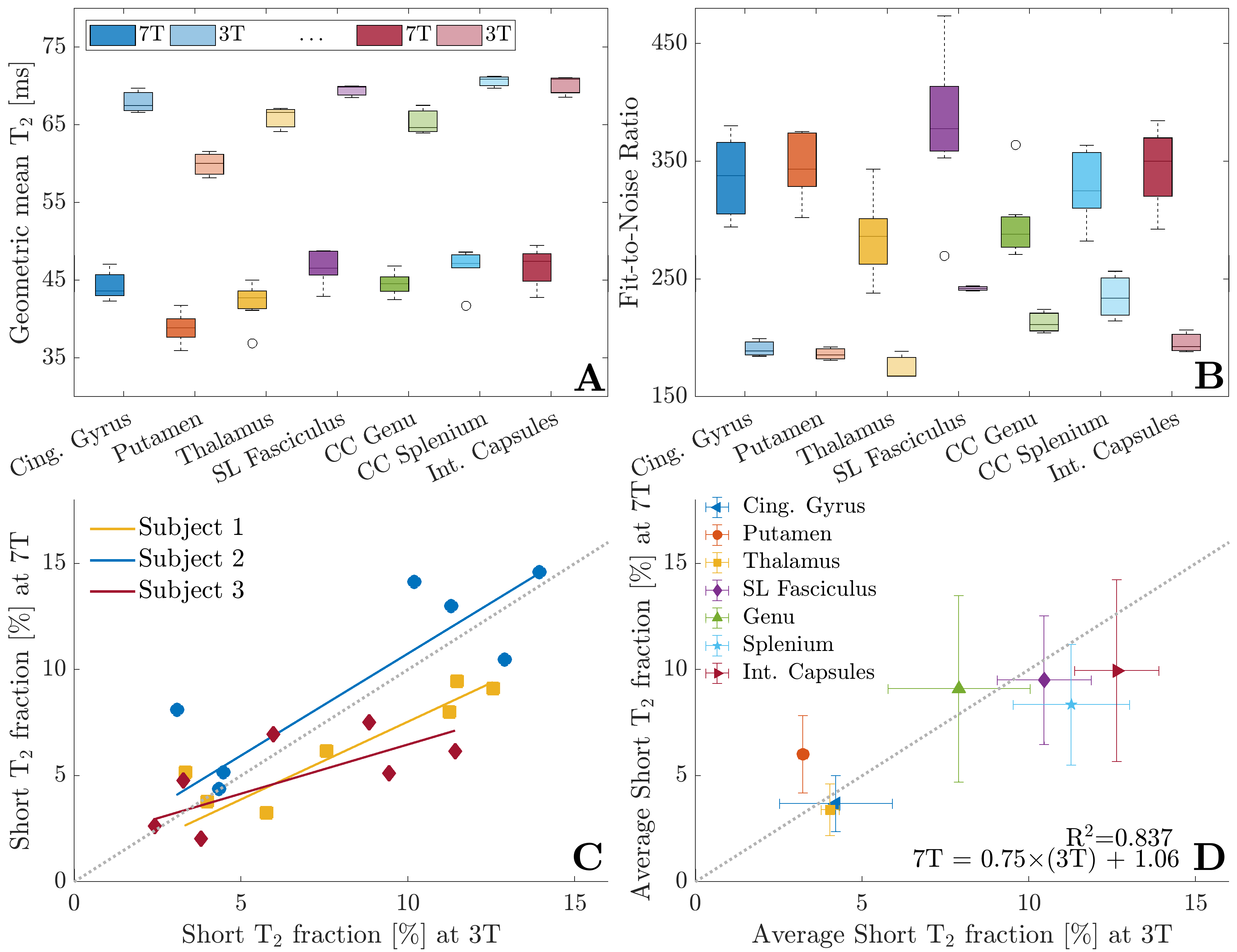}
	\phantomcaption
	\label{fig:fig6A}
	\end{subfigure}
	\begin{subfigure}[c]{0\textwidth}
	\includegraphics[width=0pt]{Upload_Fig6}
	\phantomcaption
	\label{fig:fig6B}
	\end{subfigure}
	\begin{subfigure}[c]{0\textwidth}
	\includegraphics[width=0pt]{Upload_Fig6}
	\phantomcaption
	\label{fig:fig6C}
	\end{subfigure}
	\begin{subfigure}[c]{0\textwidth}
	\includegraphics[width=0pt]{Upload_Fig6}
	\phantomcaption
	\label{fig:fig6D}
	\end{subfigure}
    \vspace{-0.9cm}
	\caption{\textbf{Average geometric mean T$_2$s of the intermediate T$_2$ peak (A) and FNR (B) at 3 T and 7 T, estimated for all ROIs. Individual (C) and ROI-based (D) comparison of the short T$_2$ fractions at the two field strengths.} The boxplots represent the distribution of all seven subjects at 7 T (darker shade), compared to the three subjects scanned at 3 T (lighter shade). (A) The intermediate mean T$_2$ was significantly shortened at 7 T in all ROIs. (B) The FNR was increased in all ROIs at 7 T by approximately 50\% compared to 3 T. (C) Comparing the short T$_2$-fractions of all ROIs within a subject yielded individual regression slopes $<$ 1, indicating some underestimation of short T$_2$ signal components at 7 T. (D) When sorting data by ROIs, it became apparent that the underestimation of the slopes is driven by higher short T$_2$ fraction in the putamen and smaller T$_2$ fractions in the splenium and internal capsules at 7 T compared to 3 T. The gray-dashed identity line is shown for visual reference.}
	\label{fig:fig6}
\end{figure}

The top row of \cref{fig:fig6} compares the geometric mean T$_2$ of the intermediate T$_2$ peak (A), and the regional FNR (B) at both field strength for all subjects. The geometric mean T$_2$ was reduced at 7 T compared to 3 T for all ROIs. At 7 T, FNR was higher by a median of 1.36 – 1.85 over the 3 T FNR and SNR was increased by a factor of 1.10 – 1.67 over 3 T SNR (not shown). The iron-rich putamen exhibited the shortest geometric mean T$_2$. In the bottom row, the median short T$_2$ fractions are compared using the previously determined T$_2$ ranges at the two field strengths, for the three “travelling brains” (C) and by ROI for the same subjects (D).

The three subject regression lines in \cref{fig:fig6C} indicate that short T$_2$ fractions tended to be slightly lower at 7 T. The individual regression slopes ranged from 0.46 -- 0.96. Comparing separate ROIs (D), we noted good agreement between the two field strengths. With the exception of high MWF WM ROIs and the iron-rich putamen, the regional mean short T$_2$ fractions were close to the unity line (gray dotted). Since high iron content can drive additional T$_2$ shortening more at 7 T and thereby increase the measured apparent short T$_2$ signal fraction \cite{36}, we excluded the putamen from the regression quoted in panel D and above. Note that the variance of all measures at 7 T was larger, likely due to a stronger influence of tissue iron and residual effects of B$_1^+$.

\section*{Discussion}

We demonstrated the feasibility of in vivo multi-component T$_2$ mapping at 7 T, building on the combination of multi-echo GraSE with EPG-modelling, as established at 3 T \cite{18}. T$_2$ distributions of various brain structures at 7 T were computed and compared to 3 T data acquired in three “travelling brains” in order to determine T$_{\text{2,min}}$ and T$_{\text{2,cut}}$ for estimating short T$_2$ fractions at 7 T. The cohort size was limited due to the geographical distance between the participating labs. Our observations are intended to serve as a starting point for further refinement of multi-echo T$_2$ measurements at 7 T.

\subsection*{7 T sequence implementation}
To ensure comparability between the multi-echo GraSE protocols at 3 T and 7 T, only parameters relevant for safety and/or measurement time were adjusted. Notably, TR of the 32-echo GraSE sequence \cite{18} needed to be prolonged at 7 T. In order to reduce scan time, additional acceleration in the head-feet direction was rendered possible by the smaller spatial coverage of the 7 T transmit coil and because the receive coils effectively suppress signal from inferior regions. In addition to SENSE acceleration, greater EPI acceleration might be of interest at 7 T. When a shorter $\Delta$TE of 8 ms or an EPI factor of 5 were attempted, peripheral nerve stimulation (PNS) increased to about 80\% of the 1st control level and volunteers reported PNS. The specific gradient coil design may have contributed to the prevalence of PNS \cite{38}. Because this can only be alleviated by decreasing in-plane resolution, sequence adjustments were not further pursued for the sake of comparability. Using the traditional parameters ($\Delta$TE = 10 ms, EPI factor = 3), PNS was at 65\% of the 1st control level and none of the subjects reported PNS or a feeling of warmth during GraSE. 

At 7 T, the WM T$_1$ is approximately 50\% longer \cite{39,40,41} than at 3 T \cite{42,43}, thus, extending TR reduced T$_1$-weighting at 7 T (\cref{fig:fig1}). Lower short T$_2$ fractions have been reported with longer TR \cite{44,45}. This observation can be linked to residual T$_1$-weighting affecting the signal due to multi-exponential T$_1$ relaxation of WM as reported by Labadie et al. \cite{46}. However, as higher Mz is available at longer TR, the estimation of the short T$_2$ component will also be more accurate as SNR is improved \cite{14,20}. Reduced noise levels are advantageous as positive noise in the first data points of the T$_2$ decay can falsely enhance short T$_2$ components \cite{14}.
At 7 T, SNR was also improved by the use of a 32-channel receive array. As indicated by the up to 85\% increase in FNR and up to 67\% increase in SNR, there is potential for further improvements regarding acceleration and spatial resolution at 7 T. The measured FNR increase was just below the predicted SNR gain of 1.91 when going from 3 T to 7 T, which is obtained from the ratio of the field strengths and considering the SENSE acceleration factors \cite{47}, ignoring differences in coil geometries and T$_1$-weighting. 
In addition to noise, B$_1^+$ inhomogeneity significantly affects T$_2$ estimations \cite{20}. We showed that the larger deviations from the ideal refocusing FA at 7 T were well accounted for by the EPG algorithm, as demonstrated by comparison to DREAM B$_1^+$-mapping (\cref{fig:fig2}). However, FA deviations led to a more rapid signal loss, especially in the temporal lobes. In turn, this may have compromised the T$_2$ characterization. High permittivity dielectric pads \cite{48} may be used to improve B$_1^+$ homogeneity, but mainly to achieve higher FA in the cerebellum. Parallel transmit techniques \cite{49,50} were not employed here as the 8-channel transmit coil was not set up at the time of data acquisition. 

\subsection*{T$_2$ distributions, selection of T$_{\text{2,min}}$ and T$_{\text{2,cut}}$ for 7 T} 
At 7 T, T$_2$ times of all water pools were shortened, so that the short T$_2$ signal was below 20 ms and intermediate water signal between 20 to 150 ms. 
At clinical field strength, MWI studies have traditionally used T$_{\text{2,min}}$ longer than TE$_1$. Recent simulation work, however, demonstrated that limiting the T$_2$ range may lead to underestimations of the MWF \cite{20}. In particular at 7 T, short T$_2$ might on average be approximately equal or shorter than 10 ms. We selected T$_{\text{2,min}}$ = 8 ms, for both 3 T and 7 T in line with recent detailed investigation of these dependencies via signal simulations \cite{20}. We found that T$_{\text{2,cut}}$ at 7 T should be at about 20 ms to exclude contamination from intermediate T$_2$ signal. This considerably less than the 40 ms reported at 3 T \cite{23}. Moreover, investigation of the 3 T distributions suggested that the equivalent threshold at 3 T should be at 30 ms. Thus, the 3 T short T$_2$ fractions obtained here may be systematically lower than previously reported. Diffusion and a larger contribution from stimulated echoes make the signal susceptible to T$_2$ shortening from static field inhomogeneities, especially in iron-rich structures. Recent histopathological work demonstrated the effect of tissue iron on the short T$_2$ component estimation \cite{36}. Given the amplification of microscopic field inhomogeneities at UHFs, brain iron in both healthy and neurodegenerative conditions deserves particular attention in future multi-component T$_2$ studies at 7 T. 
To our knowledge, this is first description of in vivo human central nervous system tissue multi-component T$_2$ properties at 7 T. Other literature pertains to animal work or fixed human brain tissue at 7 T. Both, in vivo animal work \cite{52} and post-mortem human tissue studies reported T$_{\text{2,cut}}$ = 20 ms \cite{6}, in line with our in vivo human WM findings. It should be noted that these cut-offs will change with field strengths and tissue fixation \cite{6,23} and SNR differences, e.g. achieved by signal averaging when imaging anesthetized animals, will play a role in the estimation of the short T$_2$ fraction. Our cut-off level recommendations should thus be confirmed in future studies. 

\subsection*{7 T – 3 T comparison of geometric mean T$_2$ times, FNR and short T$_2$ fraction estimates}
Literature on T$_2$ in brain tissue at 7 T is scarce due to the difficulties of measuring T$_2$ at UHFs. Mono-exponential T$_2$ studies at 7 T reported approximately 55 ms in human WM \cite{53}. This estimate is roughly in line with the 47 ms geometric mean T$_2$ of the intermediate peak in WM (\cref{fig:fig6A}). Field strength dependent changes in the relaxation rate of 1/T$_2$ have been reported to vary from linear to quadratic \cite{54,55,56}, depending on the tissue of interest. Exchange between protons in bulk water and exchangeable protons on the surface of macromolecules \cite{56} may also contribute to the field strength dependent apparent shortening of T$_2$ in central nervous system tissues \cite{53}. The short T$_2$ fractions at 7 T were close to the fractions obtained at 3 T, but slightly smaller in some WM regions. Differences between 3 T and 7 T short T$_2$ fractions were smaller than variations between individuals.  Finally, and although indiscernible on visible inspection, the iron-rich putamen presented with larger short T$_2$ components at 7 T compared to 3 T (\cref{fig:fig6D}). T$_2$ shortening due to the presence of iron appeared to be more relevant in our comparison of 7 T and 3 T than in previous studies comparing short T$_2$ fractions at 1.5 T and 3 T \cite{23}. 

\subsection*{Roadmap to myelin water imaging at UHFs}
The presented protocol transfer to 7 T and comparisons to 3 T lay the groundwork for future optimization of multi-echo T$_2$ acquisitions at 7 T. Using the TE settings established at lower field strength, the short T2 fraction maps displayed only some improvement in SNR compared to 3 T despite higher signal FNR and SNR. This may have been a consequence of artefacts from arterial inflow, due to the influence of iron and due to some short T$_2$ signal decaying before the first echo at 7 T. Further improvements of 7 T multi-echo T$_2$ acquisitions should aim at reducing TE$_1$ and $\Delta$TE. This may require sacrificing in-plane resolution in order to maintaining acceptable PNS levels, in particular for patient studies. Reducing TE may also require increasing the number of echoes. Recent MWI implementations at 3 T have used a GraSE sequence with 48 echoes at 8 ms echo spacing, albeit at voxel sizes of $1\times2\times5$ mm$^3$ \cite{59}. Although the T$_2$ relaxation characterization of human brain tissues at 7 T will likely improve at shorter TEs, SAR limits and increased PNS may hamper practical implementation. Upon establishment of suitable TE, further optimization will thus deal with how to best spend the available SNR between voxel size and acceleration.

\section*{Conclusions}
We demonstrated the feasibility of multi-echo T$_2$ GraSE measurements in vivo at 7 T for exploring the T$_2$ distributions of the brain’s WM. An accelerated T$_2$ decay was observed compared to 3 T. By variation of the T$_2$ boundaries, we established an intermediate T$_2$ range encompassing 20 -- 150 ms and a short T$_2$ peak below 20 ms. The stronger B$_1^+$-inhomogeneity at 7 T was well estimated by the EPG algorithm. Use of a multi-channel transmit coil and shortening of the echo spacing for a better definition of the short and intermediate T$_2$ water peaks will likely improve the estimation of the myelin-related, short T$_2$ component at 7 T. 


\acknow{
VW was supported by a graduate studentship award from the Multiple Sclerosis Society of Canada (EGID 2002) and is grateful for travel support from BC Children’s Hospital Research Institute through the former CFRI Research and Methodology Training Grant. ALM acknowledges support from the Natural Science and Engineering Research Council of Canada (NSERC). GH acknowledges support from the Swedish Research Council (Vetenskapsrådet, NT 2014-6193). Lund University Bioimaging Centre (LBIC) is acknowledged for providing experimental resources (equipment grant VR-RFI 829-2010-5928). AR acknowledges support from NSERC, the National Multiple Sclerosis Society and Canada Research Chairs. We wish to thank Emil Ljungberg, Shannon Kolind and John Kramer for contributing data from their studies to this project.
We are very grateful for the training data for FIRST, particularly to David Kennedy at the CMA, and also to: Christian Haselgrove, Centre for Morphometric Analysis, Harvard; Bruce Fischl, Martinos Center for Biomedical Imaging, MGH; Janis Breeze and Jean Frazier, Child and Adolescent Neuropsychiatric Research Program, Cambridge Health Alliance; Larry Seidman and Jill Goldstein, Department of Psychiatry of Harvard Medical School; Barry Kosofsky, Weill Cornell Medical Center. 
}

\showacknow{} 


\section*{References}
\bibliography{biblio}{}

\label{mylastpage}

\onecolumn

\section*{Supplemental Materials}
\setcounter{equation}{0}
\setcounter{figure}{0}
\setcounter{table}{0}
\setcounter{page}{1}
\makeatletter
\renewcommand{\thepage}{S\arabic{page}}
\renewcommand{\theequation}{S\arabic{equation}}
\renewcommand{\thetable}{S\arabic{table}}
\renewcommand{\thefigure}{S\arabic{figure}}
\renewcommand{\bibnumfmt}[1]{[S#1]}
\renewcommand{\citenumfont}[1]{S#1}

\begin{figure}[hb]
\includegraphics[width=\textwidth]{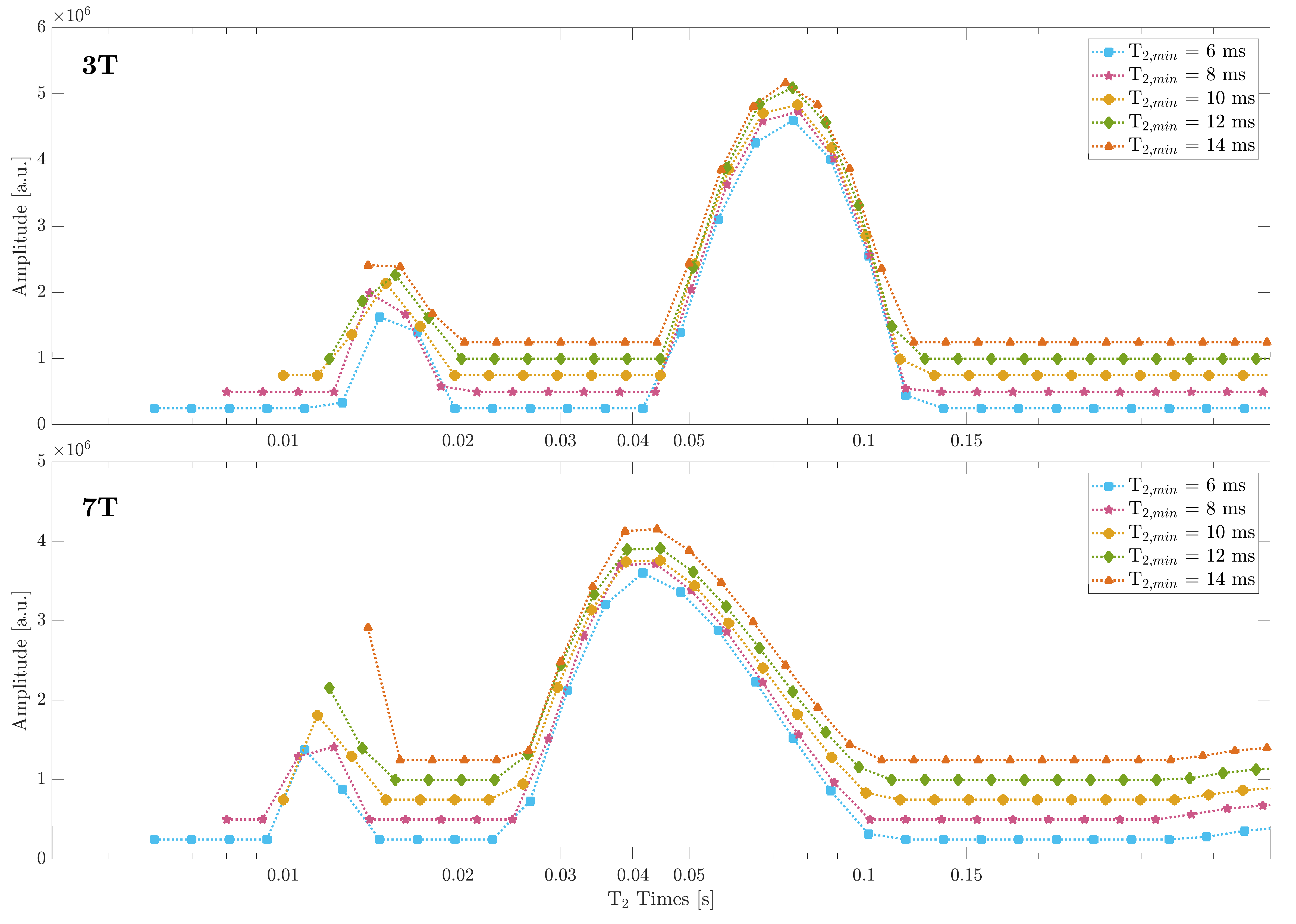}
\caption{T$_2$ distributions at 3T (top) and 7T (bottom) from a single WM voxel generated for different T$_2$ ranges using 40 logarithmically-spaced T$_2$ times. Since the distributions largely overlap, plots were stacked along $\hat{y}$ for better visibility. The choice of T$_\text{2,min}$ systematically changed the representation of the short T$_2$ peak. At longer T$_\text{2,min}$, the peak formed a slope toward the lowest available T$_2$ time, i.e. T$_\text{2,min}$. When T$_\text{2,min}$ was sufficiently small, a full peak formed around a consistent center and the estimated short fractions stabilized. Decreasing T$_\text{2,min}$ from 14 ms through 12, 10, 8 and 6 ms, the short T$_2$ fraction at 3T was 11.9\%, 12.3\%, 12.4\%, 12.4\%, 12.4\%, respectively. At 7T, the fraction amounted to 8.4\%, 8.1\%, 7.7\%, 7.5\%, 8.1\%, respectively. The intermediate water peak was unaffected by changes in T$_\text{2,min}$. Note again that T$_2$ times of both water compartments were consistently shorter at 7T than at 3T. Because the short T$_2$ signal at 3T is found at T$_2$ $>$10 ms, so that T$_2$ $>$ TE$_1$, the choice of T$_\text{2,min}$ affected the computed maps only minimally.}
\label{sfig:fig1}
\end{figure}

\begin{table}[htp!]
	\centering
	\caption{Quantitative comparison of the outcome measures with dependency on T$_{\text{2,min}}$ at 3 T and 7 T. At the same field strength, there were no significant differences between choosing T$_{\text{2,min}}$ = 8 ms and T$_{\text{2,min}}$ = 10 ms.}
	\begin{tabularx}{\linewidth}{@{\extracolsep{\fill}}>{\hsize=4.2\hsize}c|l|ccccc@{}}
		\toprule
		& T$_{2,\text{min}}$ & \makecell{Short T$_{2}$\\ fraction\\ {[\%]}} & \makecell{Geom.\  mean\\ short T$_{2}$\\ peak [ms]} & \makecell{Geom.\  mean\\ intm.\ T$_{2}$\\ peak [ms]} & \makecell{EPG-FA \\ estimation\\ {[$^{\circ}$]}} & \makecell{Fit-to-noise\\ ratio\\ (FNR)} \\
		\midrule
		\multirowcell{4}{ROI 1} & 7T$-$8ms & 3.27$\pm$3.5 & 11.2$\pm$3.2 & 46.4$\pm$3.4 & 159.0$\pm$13.0 & 372.0$\pm$98.6 \\
		& 7T$-$10ms & 3.22$\pm$3.5 & 12.4$\pm$2.9 & 46.5$\pm$3.4 & 159.0$\pm$13.0 & 369.8$\pm$96.9 \\
		& 3T$-$8ms & 5.76$\pm$4.5 & 11.0$\pm$5.1 & 67.5$\pm$5.5 & 174.9$\pm$4.3 & 184.4$\pm$50.0 \\
		& 3T$-$10ms & 5.22$\pm$4.0 & 13.0$\pm$5.4 & 67.6$\pm$5.6 & 175.0$\pm$4.3 & 182.3$\pm$49.9 \\
		\midrule
		\multirowcell{4}{ROI 2} & 7T$-$8ms & 5.17$\pm$6.0 & 11.1$\pm$3.9 & 38.8$\pm$4.4 & 174.2$\pm$5.4 & 372.3$\pm$101.1 \\
		& 7T$-$10ms & 5.34$\pm$6.0 & 12.8$\pm$3.7 & 39.0$\pm$4.4 & 174.4$\pm$5.5 & 367.9$\pm$98.8 \\
		& 3T$-$8ms & 3.33$\pm$4.4 & 12.7$\pm$6.9 & 60.0$\pm$5.7 & 165.4$\pm$7.3 & 185.1$\pm$35.9 \\
		& 3T$-$10ms & 3.04$\pm$4.0 & 14.8$\pm$7.0 & 60.1$\pm$5.7 & 165.5$\pm$7.2 & 183.4$\pm$34.9 \\
		\midrule
		\multirowcell{4}{ROI 3} & 7T$-$8ms & 3.78$\pm$4.9 & 11.4$\pm$3.6 & 41.9$\pm$5.0 & 131.1$\pm$12.3 & 276.2$\pm$80.4 \\
		& 7T$-$10ms & 3.74$\pm$4.8 & 13.0$\pm$3.5 & 42.0$\pm$5.1 & 131.2$\pm$12.3 & 274.6$\pm$79.7 \\
		& 3T$-$8ms & 3.98$\pm$4.6 & 11.8$\pm$5.5 & 64.1$\pm$5.7 & 159.5$\pm$7.5 & 167.0$\pm$42.0 \\
		& 3T$-$10ms & 3.61$\pm$4.3 & 13.5$\pm$5.5 & 64.2$\pm$5.8 & 159.7$\pm$7.5 & 165.5$\pm$41.2\\
		\midrule
		\multirowcell{4}{ROI 4} & 7T$-$8ms & 8.02$\pm$5.0 & 12.4$\pm$3.6 & 48.8$\pm$3.1 & 164.1$\pm$12.0 & 377.4$\pm$144.9 \\
		& 7T$-$10ms & 8.33$\pm$5.0 & 13.6$\pm$3.4 & 49.1$\pm$3.2 & 164.2$\pm$12.0 & 373.4$\pm$142.4 \\
		& 3T$-$8ms & 11.2$\pm$4.0 & 10.7$\pm$3.5 & 68.5$\pm$2.9 & 176.5$\pm$4.0 & 243.9$\pm$44.7 \\
		& 3T$-$10ms & 10.6$\pm$3.7 & 12.0$\pm$3.1 & 68.8$\pm$2.9 & 176.7$\pm$3.9 & 239.6$\pm$44.3 \\
		\midrule
		\multirowcell{4}{ROI 5} & 7T$-$8ms & 6.18$\pm$7.4 & 11.5$\pm$3.3 & 45.5$\pm$3.6 & 159.2$\pm$12.4 & 287.7$\pm$106.1 \\
		& 7T$-$10ms & 6.15$\pm$7.4 & 12.6$\pm$3.1 & 45.7$\pm$3.7 & 159.3$\pm$12.4 & 285.2$\pm$104.2 \\
		& 3T$-$8ms & 7.56$\pm$5.7 & 14.1$\pm$6.6 & 63.9$\pm$3.8 & 164.3$\pm$9.2 & 223.9$\pm$49.4 \\
		& 3T$-$10ms & 7.32$\pm$5.5 & 15.6$\pm$6.6 & 64.2$\pm$3.9 & 164.5$\pm$9.1 & 221.5$\pm$48.9 \\
		\midrule
		\multirowcell{4}{ROI 6} & 7T$-$8ms & 9.46$\pm$6.3 & 11.4$\pm$3.1 & 48.6$\pm$3.2 & 146.8$\pm$18.1 & 318.7$\pm$90.8 \\
		& 7T$-$10ms & 9.47$\pm$6.2 & 12.5$\pm$2.8 & 48.9$\pm$3.3 & 146.9$\pm$18.1 & 315.2$\pm$89.3 \\
		& 3T$-$8ms & 11.5$\pm$5.2 & 13.5$\pm$5.3 & 69.7$\pm$3.4 & 172.5$\pm$7.2 & 256.3$\pm$53.9 \\
		& 3T$-$10ms & 11.2$\pm$5.0 & 14.4$\pm$5.1 & 70.0$\pm$3.4 & 172.5$\pm$7.0 & 253.5$\pm$53.1\\
		\midrule
		\multirowcell{4}{ROI 7} & 7T$-$8ms & 9.12$\pm$7.2 & 13.3$\pm$3.7 & 48.2$\pm$6.7 & 160.8$\pm$12.8 & 346.4$\pm$81.6\\
		& 7T$-$10ms & 9.86$\pm$7.4 & 14.5$\pm$3.4 & 48.7$\pm$6.9 & 161.0$\pm$12.9 & 343.1$\pm$79.7\\
		& 3T$-$8ms & 12.6$\pm$7.0 & 16.8$\pm$6.2 & 71.0$\pm$8.4 & 166.2$\pm$8.1 & 192.6$\pm$39.8\\
		& 3T$-$10ms & 12.6$\pm$6.9 & 17.9$\pm$5.9 & 71.7$\pm$8.7 & 166.4$\pm$7.9 & 190.7$\pm$38.9\\
		\bottomrule
	\end{tabularx}
	\label{stab:tab1}
\end{table}

\clearpage

\end{document}